\documentclass[twocolumn,pre,amsmath,floatfix]{revtex4}
\usepackage{graphicx,color}
\usepackage{pstricks,pst-node,pst-text,pst-3d}

\newtheorem{exercise}{Exercise}[section] 
\newcommand{\dav}[2][{ }]{{\color{blue}{[}}#2{\color{blue}{]^{{\color{black}{#1}}}_{\rm av}}}}

\newif\ifpdf 
  \ifx\pdfoutput\undefined \pdffalse 
     \else 
       \pdfoutput=1 % we are running PDFLaTeX 
        \pdftrue 
 \fi
\ifpdf 
    \newcommand{\exten}{jpg}
\else 
    \newcommand{\exten}{eps}
\fi

\newcommand{\figone}{%
\begin{figure}[h]%[htbp]
   \centering
   \includegraphics[width=2.in]{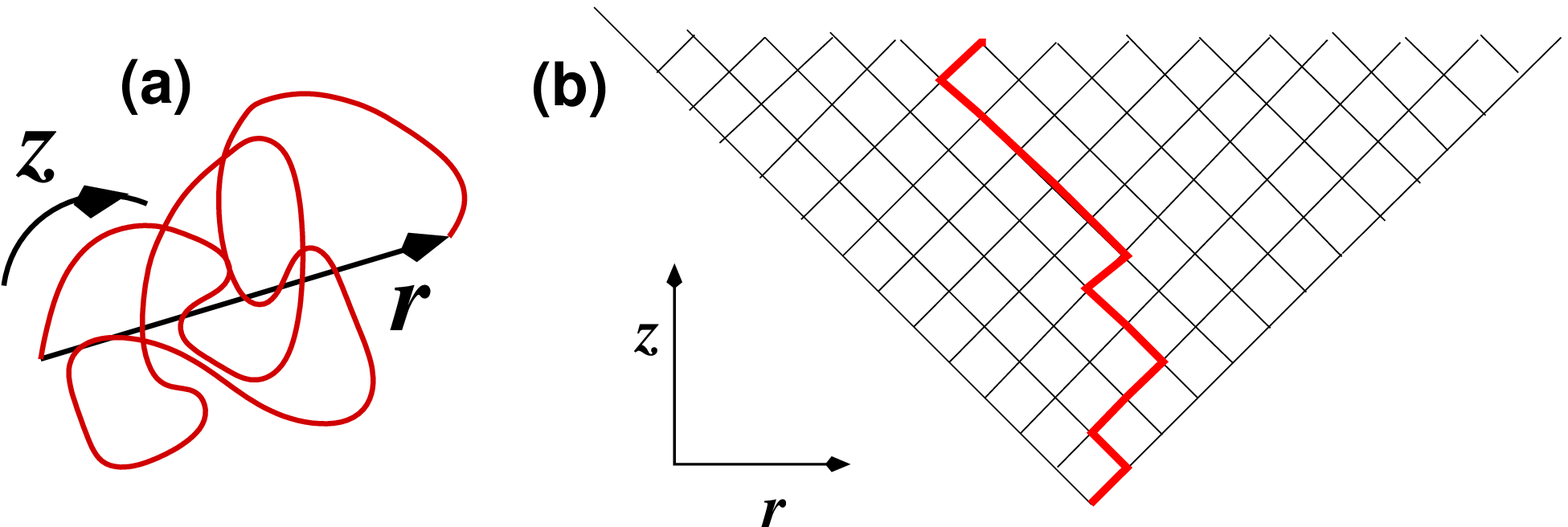}
   \caption{(a) A random walk in $d$ dimensions with $z$ as the
     variable along the contour of the polymer i.e. giving the
     location of the monomers. (b) Directed polymer on a square
     lattice.  A polymer as of (a) can be drawn in $d+1$ dimensions.
     This is like a path of a quantum particle in nonrelativistic
     quantum mechanics.} 
   \label{fig:1}
 \end{figure}
}
\newcommand{\figbox}{%
\begin{figure}[h]%[htbp]
   \centering
   \includegraphics[width=1.5in]{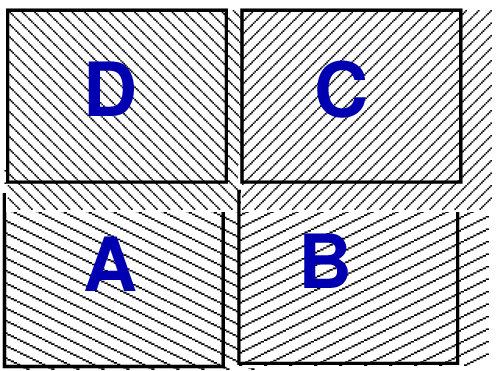}
   \caption{Adding blocks to build a larger system. } 
   \label{fig:box}
 \end{figure}
}

\newcommand{\fignN}{%
\begin{figure}[h]%[htbp]
   \centering
   \includegraphics[width=1.5in]{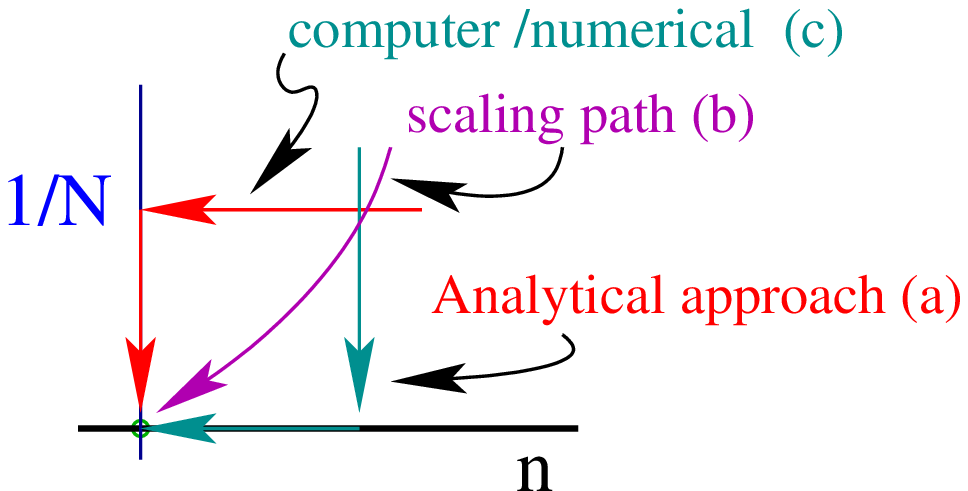}
   \caption{ Paths for replica approach } 
   \label{fig:nN}
 \end{figure}
}
\newcommand{\figreu}{%
\begin{figure}[htbp]
   \centering
   \includegraphics[width=1.in]{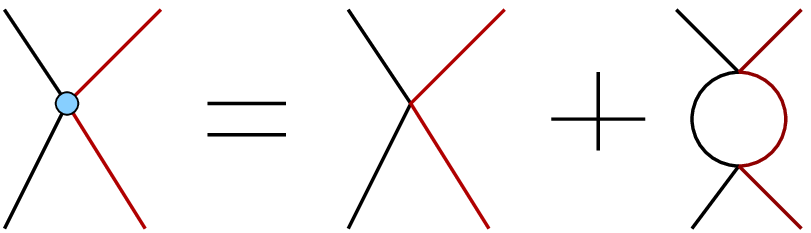}
   \caption{ Renormalization of the interaction. The polymers are
     represented by the two lines and an intersection represents an interaction. } 
   \label{fig:ufp}
 \end{figure}
}
\newcommand{\figfp}{%
\begin{figure}[htbp]
   \centering
   \includegraphics[width=2.in]{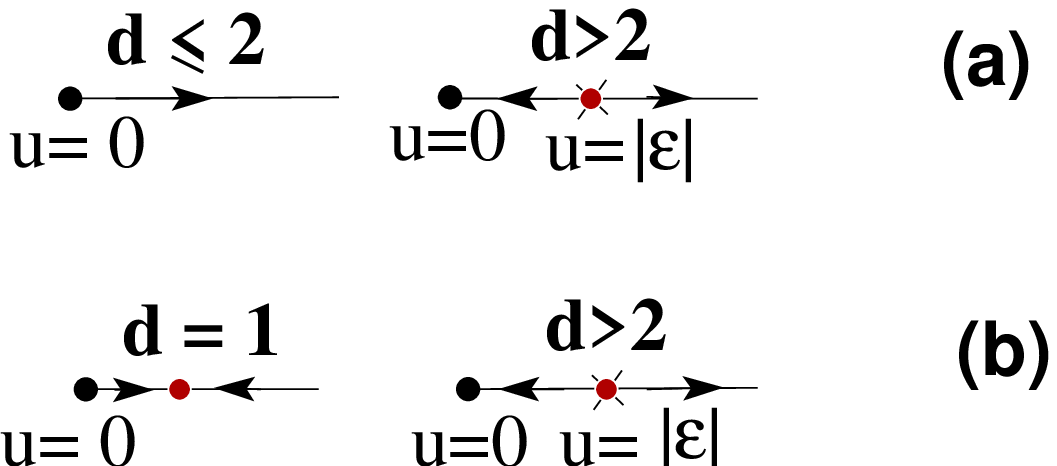}
   \caption{ RG Fixed points for $u$. (a) Based on the second moment
     of the partition function. (b) Based on the KPZ equation.
     Arrows show the flow of $u$. $\epsilon=2-d$.} 
   \label{fig:fp}
 \end{figure}
}
\newcommand{\figriv}{%
\begin{figure}[htbp]
   \centering
   \includegraphics[width=1.5in]{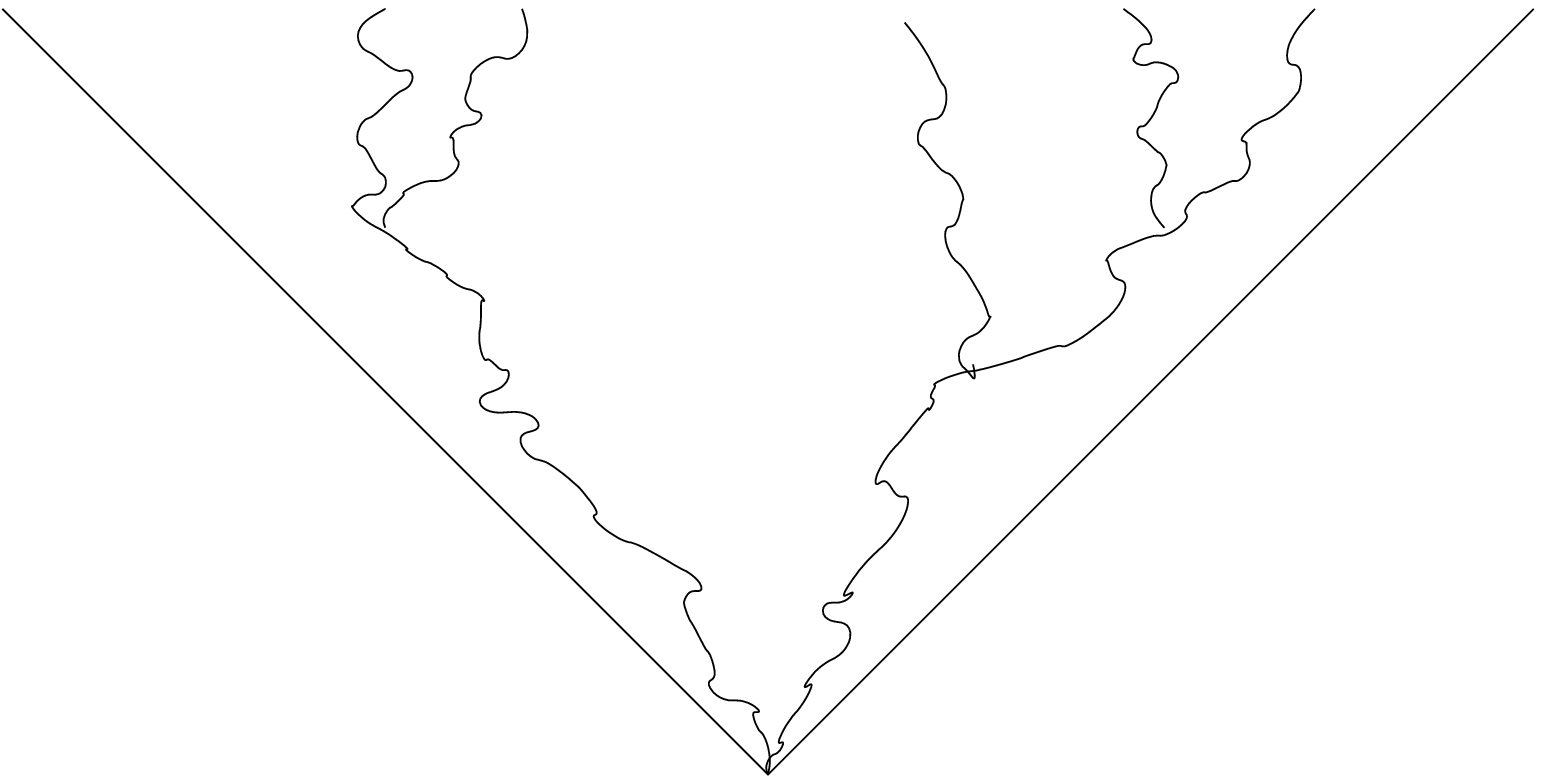}
   \caption{ Various paths for various locations of the end point.
                   } 
   \label{fig:river}
 \end{figure}
}
\newcommand{\figblob}{%
\begin{figure}[htbp]
   \centering
   \includegraphics[width=1.5in]{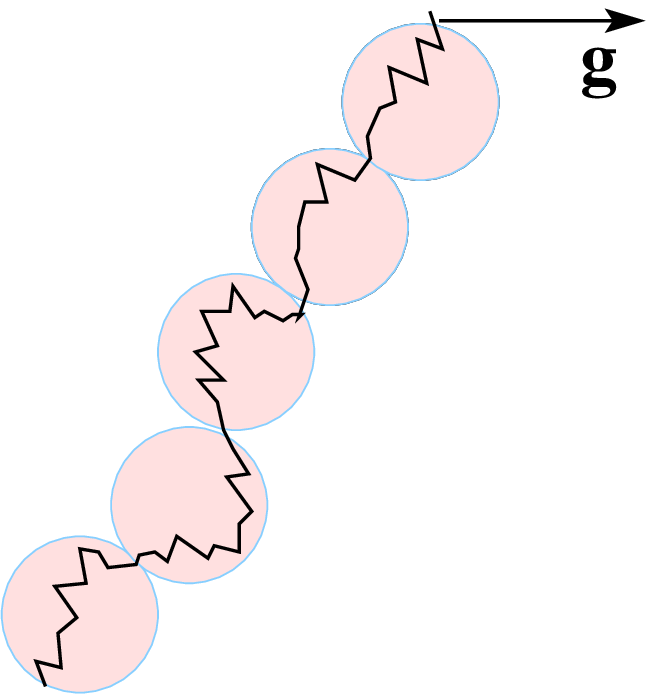}
   \caption{ A blob picture of the polymer under a force.  Though
     drawn as sphere, the $z$-direction is elongated with isotropy in
     the transverse direction.} 
   \label{fig:blob}
 \end{figure}
}

\begin{document}

\title{Directed polymer in a random medium - an introduction}
\author{Somendra M. Bhattacharjee}
\affiliation{Institute of Physics, Bhubaneswar 751 005, India\\ email:
somen@iopb.res.in }
% Institute of Physics, Bhubaneswar 751 005, India
%email: somen@iopb.res.in }
\begin{abstract}
  This is a set of introductory lectures on the behaviour of a
  directed polymer in a random medium.  Both the intuitive picture
  that helps in developing an understanding and systematic approaches
  for quantitative studies are discussed.
\end{abstract}
\date{\today}   
\maketitle

%\tableofcontents

\begin{center}
  Contents
\end{center}
\noindent{\bf {I} Directed polymers}\\
\hspace*{0.5cm} {A} Why bother?  \\
\noindent{\bf {II} Hamiltonian and Randomness}\\
\hspace*{0.5cm}  {A} Pure case\\
\hspace*{0.5cm} {B} Let's put randomness\\
\hspace*{1.cm} {1} Hamiltonian\\
\hspace*{1.cm} {2} Partition function\\
\hspace*{1.cm} {3} Averages\\
\hspace*{0.5cm}  {C} What to look for?\\
\hspace*{0.5cm}  {D} Size: disorder and thermal correlation\\
\hspace*{1.cm}  {1} Free energy fluctuation\\
\hspace*{1.cm}  {2} Our Aim\\
 \hspace*{1.cm} {3} Digression: Self-averaging\\
\noindent{\bf  {III} Q.: Is disorder relevant?}\\
\hspace*{0.5cm}  {A} Annealed average: low temperature problem\\
\hspace*{0.5cm}  {B} Moments of $Z$\\
\hspace*{1.cm}  {1} Hamiltonian for moments\\
\hspace*{1.cm}  {2} Bound state: two polymer problem\\
\hspace*{0.5cm}  {C} RG approach\\
\hspace*{1.cm}  {1} Expansion in potential\\
\hspace*{1.cm} {2} Reunion\\
\hspace*{1.cm} {3} Divergences\\
\hspace*{1.cm}  {4} RG flows\\
\hspace*{1.cm}  {5} Relevant, irrelevant and marginal\\
\hspace*{1.cm}  {6} Fixed points: strong vs weak disorder\\
 \hspace*{0.5cm} {D} { \bf Ans.: Yes, it is!}\\
\noindent{\bf {IV} Quantitative Results on exponents}\\
\hspace*{0.5cm}  {A} Replica\\
 \hspace*{1.cm} {1} $n\rightarrow 0$ and $N\rightarrow \infty $\\
 \hspace*{0.5cm} {B} Bethe ansatz\\
 {1} Flory approach\\
\noindent{\bf{V} Can we handle free energy?}\\
\hspace*{0.5cm}  {A} Free energy and the KPZ equation\\
\hspace*{0.5cm}  {B} Thermal correlation and an exact result\\
\hspace*{1.cm}  {1} Exact result on response: pure like\\
\hspace*{0.5cm} {C} Free energy of extension: pure like\\
\hspace*{0.5cm}  {D} RG of the KPZ equation\\
\hspace*{1.cm}  {1} Scale transformation and an important relation\\
\noindent{\bf {VI} Virtual reality: Intuitive picture}\\
 \hspace*{0.5cm} {A} Rare events\\
 \hspace*{0.5cm} {B} Probability distribution\\
\hspace*{1.cm}  {1} Response to a force\\
 \hspace*{1.cm} {2} Scaling approach\\
 \hspace*{1.cm} {3} And so the probability distribution is\\
 \hspace*{0.5cm} {C} Confinement energy\\
\noindent{\bf{VII} Overlap}\\
\noindent{\bf {VIII} Summary and open problems}\\
\hspace*{0.5cm}  {}Acknowledgments\\
\hspace*{0.5cm}  {}References\\

\section{Directed polymers}
A physicist's polymer is a long string like object, devoid of any
specific inner structure.  It can be treated as a random walk on a
lattice or as an elastic string in continuum.  The lattice problem is
well-defined though in many situations the continuum version is easier
to treat.  For the lattice problem, the sites visited are the
monomers with the steps as bonds.  The monomers are indexed by a
coordinate $z$ which becomes a continuous variable in the continuum
model.   See Fig. \ref{fig:1}.

An important quantity for a polymer is its size or the spatial extent
as the length $N$ becomes large.  For a
translationally invariant system with one end ($z=0$) fixed at origin,
the average position at $z=N$ is zero but the size is given by the rms value
\begin{equation}
  \label{eq:9}
  \langle {\bf  r}_N\rangle=0,\quad \langle r^2_N\rangle ^{1/2}\sim 
  N^{\nu},
\end{equation}
with $\nu=1/2$, for the free case.  In
general the size exponent $\nu$ defines the polymer universality class
and it depends only on a few basic elements of the polymer.  In addition
to this geometric property, the usual thermodynamic quantities, e.g.
free energy (or energy at $T=0$), are also important, especially if
one wants to study phase transitions.

\figone

We consider the problem of a polymer where each monomer sees a
different, independent, identically distributed random potentials.
Geometrically this can be achieved if the monomers live in separate
spaces.  One way to get that is to consider the polymer to be a $d+1$
dimensional string with the monomers in $d$ dimensional planes but
connected together in the extra dimension.  As shown in Fig.
\ref{fig:1}, this is a polymer which is directed in one particular
direction.  Hence the name directed polymer.  

For a directed polymer, the size would now refer to the  size in the
transverse $d$-directions and so  Eq. (\ref{eq:9}) refers to the
transverse size  as the length in the special $z$-direction increases

\subsection{Why bother?}
The major issue for the random problem is to see if the universality
class of the pure polymer changes to a new one in presence of disorder
and if any new phase transition can take place.

Because of randomness (or disorder), any quantity of interest has to
be a stochastic variable (realization dependent).  Therefore, an
additional disorder averaging needs to be done over and above the
usual thermal averaging for each realization.  For thermal
fluctuations, there is generally no need to go beyond second moments
or cumulants (related to various response functions like the specific
heat, susceptibility etc.)  but that cannot be said offhand for the
disorder averaging.  The randomness is put in by hand and is not
thermalizable.  There are possibilities of rare events for which it is
necessary to distinguish between the average value and the typical
(i.e. most probable) value.  In such situations higher moments of the
quantity concerned become important.  It is this aspect that makes
disorder problems interesting and difficult.

\rule{2cm}{0.5mm}
%\begin{quotation}
{\bf Example:} {\it Typical vs. average}\\
Consider a random variable $x$ that takes two values
\begin{subequations}
\begin{eqnarray}
\label{eq:rar1}
X_1&=& e^{\alpha \sqrt{N}} \ {\rm and} \ X_2= e^{\beta N}, \beta >1,\\
\lefteqn{\mathrm{with\   probabilities}\hspace{2cm}}\nonumber\\
p_1&=&1-e^{-N}, \ {\rm and} \ p_2=e^{-N}.
\label{eq:rar2}
\end{eqnarray}
\end{subequations}
In the limit $N\rightarrow \infty$, the average value $\dav{x}
\rightarrow e^{(\beta-1)N}$  while the typical or most probable value
is $x=X_1$ with probability $1$.   On the other hand $\dav{\ln x}
\rightarrow \alpha \sqrt{N}$ in the same limit, showing that $\dav{\ln x}$
is determined by the typical value of the variable while the moments
are controlled by the rare events.   Note that this peculiarity
disappears if $x$ has a smooth probability distribution in the sense
of  no special or rare events.
\rule{2cm}{0.5mm}
%\end{quotation}

The significance of directed polymer lies in the fact that the pure
system is very well understood and exactly solvable in all dimensions.
Polymers show critical-like  behaviour (e.g., power
laws for size) without any need of any fine tuning.  One 
hopes that the understanding gained in this topic on the role of
disorder may be useful in other situations as well.   Wistful sighs...

{\bf Background:} Familiarity with
the basic ideas of renormalization group or critical phenomena
(Ref. \cite{crit}) and polymers (Refs. \cite{degen},\cite{doi})  would
be useful.

\section{Hamiltonian and Randomness}
\label{sec:hamilt-rand}

\subsection{Pure case}
\label{sec:pure-case}
Taking the polymer as an elastic string, one may define a Hamiltonian
\begin{equation}
\label{eq:1.1}
H_0=\frac{d}{2} K \int dz \left( \frac{\partial {\bf r}}{\partial
z}\right ) ^{^2}
\end{equation}
which gives a normalized probability distribution of the position
vector ${\bf r}$ at length $z$
\begin{equation}
\label{eq:1.2}
P({\bf r},z) = \frac{1}{(2\pi z)^{d/2}} \ e^{- r^2/2z} \qquad
(Kd/k_{\rm B}T =1).
\end{equation}
For the lattice random walk, there is no ``energy''  and the elastic
Hamiltonian of Eq. \ref{eq:1.1}  just simulates the entropic effect at
non-zero temperatures. One needs to look at the lattice problem in
case one is interested in low or zero temperature bahaviour.  

For a polymer of length $N$ the probability distribution gives
\begin{equation}
\label{eq:1.3}
\langle {\bf r}\rangle=0, \quad \langle  r^2\rangle= N, \qquad (K d/k_{\rm B}T=1)
\end{equation}
so that the transverse size of the polymer is given by 
\begin{equation}
\label{eq:1.3a}
R_0 \sim\langle  r^2\rangle^{1/2}\sim N^{\nu}\ {\rm with}\ \nu=1/2.
\end{equation}
The power law growth of the size of a polymer as the length increases
in a reflection of the absence of any ``length scale'' in the
Hamiltonian.

\begin{exercise}
Put in $d$, $K$ and $T$ in
Eqs. \ref{eq:1.2}, \ref{eq:1.3}, \ref{eq:1.3a}.\\
  Get Eq. \ref{eq:1.3} from Eq. \ref{eq:1.1}.
\end{exercise}

\subsection{Let's put randomness}

\subsubsection{Hamiltonian}
\label{sec:hamiltonian}
Let us now put this polymer in a random medium.  In the lattice model
of Fig. \ref{fig:1}, each site has an independent random energy and
the total energy of the lattice polymer is the sum of the energies of
the sites visited.  In continuum, the Hamiltonian can be written as
\begin{equation}
\label{eq:2.1.1a}
H=H_0
+ \int_0^N dz\  \eta({\bf r},z)\  \delta({\bf r}(z)-{\bf r})
\end{equation}
where $\eta({\bf r},z)$ is an identical, independent gaussian distributed
random variable with zero mean and variance $\Delta>0$,
\begin{equation}
  \label{eq:3}
\dav{\eta({\bf r},z)}=0, \  \dav{\eta({\bf r},z)\ \eta({\bf r}^{\prime},z^{\prime})}
=\Delta\delta({\bf r}-{\bf r}^{\prime})\delta(z-z^{\prime}) .
\end{equation}

With this distribution of random  energies, we see $\dav{H}=H_0$
and so  the  average Hamiltonian is not of much use.

\subsubsection{Partition function}
\label{sec:partition-function}
The partition function for a polymer in a random medium or potential
is given by
\begin{equation}
\label{eq:2.1.1}
Z=\int {\cal DR} \ e^{-\beta H}.
\end{equation}
This is a symbolic notation to denote sum over all configurations and
is better treated as a continuum limit of a well-defined lattice
partition function
\begin{equation}
\label{eq:2.1.2}
Z=\sum_{\rm paths} e^{-\beta \eta({\bf r},z)}
\end{equation}
where the sum is over  all possible paths of $N$ steps starting from
${\bf r}={\bf 0}$ at $z=0$.  
and $Z_0$ is the partition function of the
free walker.   
It often helps to  define the partition function such that
 $Z(\{\eta=0\}) = 1$ to avoid problems of going to the continuum
 limit (see Eq. \ref{eq:1.2}).  This is done by dividing (or
 normalizing)  $Z$ by $Z_0=\mu^N$, $Z_0$ being the  partition function of the
free walker with $\mu$ as the connectivity constant ($=2$ for Fig. \ref{fig:1}b).

\subsubsection{Averages}
\label{sec:averages}
Because of the randomness, there is no unique partition function but
rather a probability distribution of the partition function, $P(Z)$,
is needed.  Any quantity of interest needs to be averaged over such a
distribution.  This averaging is to be called {\it sample averaging},
denoted by $\dav{...}$ (as opposed to thermal averaging, denoted by
$\langle...\rangle$).  Consequently, sample to sample fluctuations may
become important and would be a new class of quantities which are not
meaningful in the pure thermal case.  The extreme situation is the
possibility of ``rare'' events which require distinctions between
typical and average behaviour.

\subsection{What to look for?}
For the pure case ($\Delta=0$), there is no ``energy'', only
configurational entropy of the polymer.
But with $\Delta \neq 0$, there may be  one or more lowest energy
states.  What is the nature of the ground state?

\begin{exercise}
  For the lattice problem, the energy is the sum of $N
  (\rightarrow\infty)$ random energies of the sites visited. This is
  tantalizingly similar to what one needs for applying the central
  limit theorem. Is it so?
\end{exercise}
\subsection{Size: disorder and thermal correlation}
\label{sec:size:-disord-therm}
So far as the geometrical properties are concerned, we first note the
lack of translational invariance for a particular realization of
disorder and therefore $\langle {bf r}\rangle \neq 0$, but on averaging
over randomness, translational invariance will be restored
statistically and so $\dav{\langle {bf r}\rangle}$ has to be zero.  One may
therefore consider the size of the polymer by the ``disorder''
correlations
\begin{equation}
\label{eq:10}
C_{\rm dis}\equiv \dav{\langle {bf r}^2\rangle}\ {\rm or}
\  C_{\rm dis}\equiv \dav[2]{|\langle   {bf r}\rangle|}\ {\rm with} \ C_{\rm dis}\sim N^{2\nu}
\end{equation}
Furthermore,  the usual correlation function is the  thermal correlation
\begin{equation}
\label{eq:11}
C_{\rm T}\equiv \dav{\langle {bf r}^2\rangle} - \dav{\langle {bf r}\rangle^2}.
\end{equation}
These are defined at all temperatures.  Of course, the probability
distribution $P(\langle {bf r}\rangle)$ is also of importance.

\subsubsection{Free energy fluctuation}
\label{sec:free-energy-fluct}
Apart from these geometric properties one would like to know the
fluctuations in free energy at any non-zero temperature (or
ground-state energy at $T=0$).  The probability distributions for
these quantities would be the ideal things to look for.  In absence of
knowledge of this distribution, one may look for various moments and
fluctuations.

Let us define $F({\bf r},N) =-T\ln Z({\bf r},N)$, a restricted free
energy that the end at $z=N$ is at ${\bf r}$.  This free energy, being
realization dependent, is also a stochastic quantity.  We may then
define a correlation function
\begin{equation}
  \label{eq:6}
  C_{F}({\bf r},N)=\dav{(F({\bf r}+{\bf r}_0,N)-F({\bf r}_0,N))^2},
\end{equation}
which may be assumed to satisfy a scaling form
\begin{equation}
  \label{eq:7}
   C_{F}({\bf r},N)=N^{2\theta} \ {\sf  f}\left(\frac{{\bf r}}{N^{\nu}}\right).
\end{equation}
Note that the exponent $\theta$ has no meaning in the pure case where
$\theta=0$.  This is a new quantity for the disordered system. 
$P({\bf  r},z)$ of  Eq. (\ref{eq:1.2}), if used for $Z({\bf  r},z)$,
satisfies Eq. (\ref{eq:7})  with $\theta=0$ and $\nu=1/2$.

For $r\ll N^{\nu}$, $C_F$ should be a function of $r$.  Taking ${\sf
  f}(x)\sim x^p$ for $x\rightarrow 0$, we need $p=\theta/\nu$ and  get
\begin{eqnarray}
  \label{eq:8}
  C_{F}({\bf r},N)&=& r^{2\theta/\nu}, \quad {\rm for}\
  r\ll N^{\nu}\nonumber\\
&=& N^{2\theta}. \qquad {\rm for}\  r\ge N^{\nu}.
\end{eqnarray}

\subsubsection{Our Aim}
Our aim is two-fold: to get the quantitative estimates of $\nu$ and
$\theta$, and to develop an intuitive picture.  In fact, there are
quantitative changes in both the exponents but, more importantly,
those changes are consequences of a drastic change in the qualitative
behaviour of the polymers.  At the end, however, a simple picture
borrowed from pure polymers remain valid.
 
{\bf On Notation:} 
\begin{itemize}
\item To avoid proliferation of symbols, we reserve the
symbol ${\sf f}$ to denote an arbitrary  or unspecified function, not
necessarily same everywhere.
\item  {\it Sample averaging} is denoted by $\dav{...}$ 
 while {\it thermal averaging} is denoted by $\langle...\rangle$.
\item The Boltzmann constant is set, most often, to one, $k_{\rm
    B}=1$.
\item ``Disorder'' and ``randomness'' will be used interchangeably.
\end{itemize}
\begin{small}
\subsubsection{Digression: Self-averaging} 
Our interest is in large systems.  A question of building up such a
large system arises.  In the pure case, we may add blocks
systematically (see Fig. \ref{fig:box} ) to generate a unique way of
going to the limit.  For a random system, it is not so
straightforward.  For a given block A, there are many possibilities
for each of B,C,D..., well, calling for a probabilistic description.
Even the issue of thermodynamic limit has to be probabilistic in
nature; we may expect $\ln Z$ to be extensive for large size with
probability one but not necessarily for each and every realization.  A
quantity of import would be the probability distribution (sample to
sample variation) of the free energy itself, $P(\ln Z)$.  If $P(\ln
Z)$ is sharply peaked for large size, then any large sample would show
the average behaviour.  A quantity with this property is often called
{\it self-averaging}.  This may not be the case if the distribution is
broad especially in the sense discussed in Sec. 1.1.  A self-averaging
quantity has the advantage that one may study one realization of a
large enough system without any need of further disorder or sample
averaging.

\figbox

To be quantitative, let us choose a quantity $M$ which is extensive
meaning $M=N m$ where $m$ is the ``density'' or per particle value.
This is based on the additive property over subsystems $M=\sum M_i$.
For a random system we better write $M=M(N,\{Q\})$, with $\{Q\}$
representing all the random variables.  To recover thermodynamics, we
want $\dav{M}$ to be proportional to $N$ for $N\rightarrow\infty$.
Now, if it so happens that for large $N$
\begin{equation}
\label{eq:2.2.1}
M(N,\{Q\}) \rightarrow N m_d,
\end{equation}
with $m_d$ independent of the explicit random variables, then $M$ is
said to have the self averaging property.  Note that no averaging has
been done in Eq. \ref{eq:2.2.1}.  One way to guarantee this
self-averaging is to have a probability distribution
\begin{equation}
  \label{eq:23}
  P(M/N) \stackrel{N\rightarrow\infty}{\longrightarrow} 
\delta(m_d).
\end{equation}
This is equivalent to the statement that the sum over a large number
of subsystems gives the average value, something that would be
expected in case the central limit theorem (CLT) is applicable.  But
for many critical systems CLT may not be applicable.

A more practical procedure for testing self-averaging behaviour of a
quantity $X$  is to study the
fluctuations $\sigma_N^2 \equiv \dav{X^2} - {\dav[2]{X}}$ and
then check if 
\begin{equation}
\label{eq:2.2.2}
R_{X,N}=\frac{\sigma_N^2}{\dav[2]{X}} \rightarrow 0, \ {\rm as\ }
N\rightarrow \infty.
\end{equation}
   One may define more general measures like $R_{X,N}^{(p)} =
{\sigma_N^{(p)}}/{{\dav[p]{X}}}$ where ${\sigma_N^{(p)}}$ is the $p$th
cumulant.  A quantity is not self-averaging if the corresponding $R$
does not decay to zero. A very slow decay (``weakly'' self averaging)
would also signal practical difficulties.
A thumb rule would be that an extensive quantity is generally
self-averaging, others may not be.  
\end{small}

\section{Q.: Is disorder relevant?}
\label{sec:disorder-relevant}
Let us first see if disorder is at all relevant.
 
\subsection{Annealed average: low temperature problem}
The averaged Hamiltonian is of no use but what about the average
partition function $\dav{Z}$?  This is called  an annealed
average.   With a gaussian distribution for the random energy, from
Eq. (\ref{eq:2.1.2}), 
\begin{eqnarray}
\label{eq:2.1.5}
\dav{Z}=  \exp(\beta^2\Delta N/2) \exp(N\ln \mu),
\end{eqnarray}
so that $F/N = -T[ \ln \mu + \beta^2 \Delta/2]$  The entropy obtained
from this partition function ($S=-\partial F/\partial T$) by
definition has to be positive. 
This gives a limit $T\geq T_{\rm A}\equiv \sqrt{\frac{\Delta}{2\ln\mu}}$.
Annealed averaging will not work at very low temperatures.

\subsection{Moments of $Z$}
For thermodynamic quantities we know that it is the average of the
free energy that is required.  This is called \textit{quenched averaging}.

We may write 
\begin{eqnarray}
  \label{eq:2}
\dav{\ln Z} &=& \dav{\ln \{\dav{Z} + (Z-\dav{Z} )\} }\nonumber\\
&=& \ln \dav{Z} +\frac{\dav{Z^2} -\dav[2]{Z} }{2\dav[2]{Z}} 
+...
\end{eqnarray}
to show the importance of the variance of the partition function.  If
the variance remains small, in the limit $N\rightarrow\infty$, then
the polymer can be described by the average partition function which
is more or less like a pure problem. Otherwise not.

\subsubsection{Hamiltonian for moments}
\label{sec:hamiltonian-moments}
Higher order terms involve higher moments $\dav{Z^n}$ or rather
cumulants.  These moments can be written as the partition function of
an $n$-polymer problem with an extra interaction induced by the
disorder.  Starting from $H$ as given by Eq. (\ref{eq:2.1.1a}), and
the gaussian distribution of Eq. (\ref{eq:3}), we have
\begin{eqnarray}
  \label{eq:4}
  \dav{Z^n} &=& \int  \left ( \int {\cal DR}\ e^{-\beta
      H}\right)^{^{\scriptstyle n}} 
     P(\eta) {\cal D}\eta \nonumber\\
  &=& \int {\cal DR}_1\ {\cal DR}_2...{\cal DR}_n 
      e^{-\beta \sum H_{0i}}\times\nonumber\\
  &&\qquad  {\psframebox[linecolor=red]
                     { \dav{e^{-\beta\int \sum \delta({\bf r}_i(z)-{\bf
                             r})\eta({\bf r},z)}}
                      }
                     }\rnode{NA}{} \nonumber\\[0.25cm]
  &=&\int {\cal DR}_1\ {\cal DR}_2...{\cal DR}_n 
                  e^{-\beta \sum H_{0i}} \times\nonumber\\
  &&\qquad {\psshadowbox[linecolor=red]{
                        \exp\left(  \beta^2 \int \Delta 
                          \sum_{i<j} \delta({\bf r}_i(z)-\rnode{ND}{}{\bf r}_j(z))\right )
                       }
                  }\rnode{NB}{}
     {\nccurve[linecolor=red,angleA=0,angleB=0,nodesep=3pt]{<->}{NB}{NA}}
       \nonumber\\
  &\equiv&\int {\cal DR}_1\ {\cal DR}_2...{\cal DR}_n \ e^{-\beta  H_n},\nonumber\\
%\end{eqnarray}
%\begin{eqnarray}
 {\rm where\ }  H_n&=& \frac{dK}{2} \int_0^N dz \sum_i^n\left( \frac{\partial {\bf
        r}_i}{\partial z}\right ) ^{^2} \nonumber\\
&&\quad         {\psframebox[linecolor=red]{
                              {\color{red}{-}} \beta \Delta  \int_0^N d z \ 
                                \sum_{i<j} \delta({\bf
                                  r}_i(z)-\rnode{NC}{}{\bf r}_j(z))
                         }
                      }.
 \label{eq:5}
\end{eqnarray}
{\nccurve[linecolor=red,angleA=270,angleB=90,
  nodesepA=12pt,nodesepB=12pt]{->}{ND}{NC}} 
This particular form can be understood in terms of two polymers.
These two polymers start from the same point and do their random walk
as they take further steps.  If there is a site which is energetically
favourable, both the polymers would like to be there.  The effect is
like an attractive interaction between the two polymers - an
interaction induced by the randomness.

\subsubsection{Bound state: two polymer problem}
\label{sec:bound-state:-2}
For the second moment, we have a two polymer problem.  The analogy
with quantum mechanics tells us that for $d<2$, any binding potential
can form a bound state but a critical strength is required for a bound
state for $d>2$.  In the polymer language, this means that any small
disorder will change the behaviour of the free (pure) chain for $d<2$
({\it disorder is always relevant}) but for $d>d_c=2$, if $\beta
\Delta <(\beta\Delta)_c$, the chain remains pure-like ({\it disorder
  is irrelevant}).  Actually in higher dimensions ($d>2$) the
delta-function potential needs to be regularized appropriately (e.g.,
by a ``spherical'' well).  Such cases are better treated by
renormalization group (RG) which also helps in making the definitions
of relevance/irrelevance more precise.  We discuss this below.

In short, the second term (fluctuation in partition function) in the
expansion of Eq. (\ref{eq:2}) cannot be ignored if $d<2$ or if
$\beta\Delta$ is sufficiently large for $d>2$.  This signals a
disorder dominated phase for all disorders in low dimensions or at low
temperatures (weak disorders) in higher dimensions.

\subsection{RG approach}
\label{sec:zrg}
It is instructive to develop a renormalization group  approach for
the two polymer problem.

\subsubsection{Expansion in potential}
\label{sec:expansion-potential}
We do an expansion in the interaction potential and just look at the
first contributing term.  Full series can of course be treated
exactly.  On averaging, the first order terms drop out, yielding
\begin{eqnarray}
\label{eq:2.1.3a}
\dav{Z^2} - \dav[2]{Z}&=& 
\int {\cal DR}_1\ \int {\cal DR}_2 e^{-\beta H_{0,1}}
e^{-\beta H_{0,2}} \times\nonumber\\ 
&&\quad \int dz \beta^2 \Delta \delta(r_1(z)-r_2(z)) + ... .
\end{eqnarray}
This is the first order term if Eq. (\ref{eq:5}) is used. 
A diagrammatic representation is often helpful in book-keeping as
shown in Fig. \ref{fig:ufp}  but we avoid the details here.

\subsubsection{Reunion}
\label{sec:reunion}
For this two polymer problem, the interaction contributes whenever
there is a meeting or reunion of the two polymers at a site.  At the
order we are considering, there is only one reunion but this reunion
can take place anywhere along the chain and anywhere in the transverse
direction.

\figreu

The probability that two walkers starting from origin would
meet at ${\bf r}$  at $z$ is given by $P^2({\bf r},z)$ (Eq. (\ref{eq:1.2}))
so that a reunion anywhere is given by a space integral of this
probability which gives 
\begin{equation}
\label{eq:1.5a}
{\cal R}_z  \equiv \int d{\bf r} P^2({\bf r},z) = (4\pi z)^{-\Psi}, 
\ {\rm with} \ \Psi=d/2. 
\end{equation}
This exponent $\Psi$ will be called the reunion exponent.  The
occurrence of a power law is again to be noted.  The eventual
renormalization group approach hinges on this power law behaviour.

\subsubsection{Divergences}
\label{sec:divergences}
The contribution in Eq. (\ref{eq:2.1.3a}) apart from some constants
involve an integral over the reunion 
behaviour given in Eq. \ref{eq:1.5a}.   This integral in the limit
$N\rightarrow\infty$ is 
\begin{eqnarray}
\label{eq:2.1.4}
\int_a^N dz \frac{1}{z^{d/2}} &\sim& {a^{1-d/2}} \quad   {\rm for}\ d>2\\
&\sim& {N^{1-d/2}}\quad   {\rm for}\ d<2.
\end{eqnarray}
For a finite cut-off, as is usually the case, the integral is finite
for $d>2$, and therefore $\dav{(Z - \dav{Z})^2} \approx
O(\beta^2\Delta)$.

This however is not the case for $d<2$ with $d=2$ as a borderline case.
The problem we face here is ideal for a renormalization group approach.

\subsubsection{RG flows}
\label{sec:rg-flows}
Let us introduce an arbitrary length scale $L$ in the transverse
direction and define a dimensionless ``running'' coupling constant
\begin{equation}
  \label{eq:1}
  u(L)= (\beta^3 K \Delta) L^{\epsilon}, \quad \epsilon=2-d
\end{equation}
and study the RG flow of the coupling constant as the scale $L$ is
changed.   This takes care of reunions at small scales to define the
effective coupling on a longer scale (with rescaling).  The flow equation is
\begin{equation}
\label{eq:uflow}
L\frac{du}{dL} = (2-d)u + u^2  
\end{equation} 
The magnitude of the coefficient of the $u^2$ term is not very crucial
because at this order, this coefficient can be absorbed in the
definition of the $u$ itself.  What matters is the sign of the $u^2$
term.
\begin{exercise}
  By change of variables in Eq. (\ref{eq:5}), show that $u(L)$ as
  defined in Eq. (\ref{eq:1}) is the correct dimensionless variable.
\end{exercise}

\subsubsection{Relevant, irrelevant and marginal}
\label{sec:relev-irrel-marg}
A coupling that grows (decays) with length scale is called a {\it
  relevant (irrelevant)} term because at long scales the contribution
of this quantity cannot (can) be ignored.  A coupling that does not
change with scale is called a {\it marginal} variable.

For example at the pure fixed point ($u=0$), disorder is relevant (see
Eq. (\ref{eq:1}) ) for $d<2$ but becomes irrelevant for $d>2$, while it
is marginal at $d=2$.  Marginal variables are important because the
renormalization procedure may add higher order corrections, like the
$u^2$ term in Eq. (\ref{eq:uflow}), which then determines the growth
or decay of the term.  Here we see that the disorder is a {\it
  marginally relevant variable}, eventually increasing with length (at
$d=2$).  A
marginally relevant variable introduces a new phase transition (or
fixed point) in higher dimensions.  The emergence of the new fixed
point for $d>2$ signals a phase transition.

\subsubsection{Fixed points: strong vs weak disorder}
\label{sec:fixed-points}
Since $u$ comes from the variance of the distribution, it cannot be
negative.  We therefore need to concentrate only on the $u\geq 0$ part
with the initial condition of $u(L=a)=u_0$.  What we see that for
$d<2$, the flow on the positive axis goes to infinity indicating a
strong disorder phase for any amount of disorder provided we look at
long enough length scale.  An estimate of this length scale may be
obtained from the nature of divergence for a given $u_0$.  An
integration of the flow equation gives $L\sim u_0^{1/(2-d)}$ ($d<2$), a
crossover length beyond which the effect of the disorder is
appreciable.

\figfp

For $d>2$, there is a fixed point at $u^*=|\epsilon|$ where
$\epsilon=2-d$.  For $u<u^*$, the disorder strength goes to zero and
one recovers a ``pure''-like behaviour.  This is a \textit{weak
  disorder} limit.  But, if $u>u^*$ the disorder is relevant.  

\subsection{{\bf Ans.: Yes, it is!}}
Based on the fixed point analysis, we conclude, as already mentioned,
that disorder can be relevant depending on the dimensions we are in
(i.e. the value of $d$) and temperature or strength of disorder.  In particular,
follows:
\begin{enumerate}
\item A disorder-dominated or strong disorder phase for all
  temperatures for $d\le 2$.
\item A disorder dominated or strong disorder phase at low
  temperatures for $d>2$.
\end{enumerate}

For $d>2$, one sees a phase transition by changing the strength of the disorder or
equivalently temperature for a given $\Delta$.  This is an example of
a phase transition induced by disorder which cannot exist in a pure
case.
The flow equation around the
fixed point for $d>2$ shows that one may define a ``length-scale''
associated with the critical point that diverges as $\xi \sim
|u-u^*|^{-\zeta}$ with $\zeta=1/|2-d|$.   In the weak disorder phase
where the disorder is irrelevant, $\dav{\ln Z}  \approx \ln \dav{Z}$,
and therefore one may put a bound on the transition temperature $T_c$
for a lattice model as $T_c\geq T_A$ as defined below Eq. (\ref{eq:2.1.5}).

In absence of any fixed point for the strong disorder phase in this
approach, no further quantitatie results can be obtained about the
phase itself.

\section{Quantitative Results on exponents}
A strong disorder phase has characteristic sample fluctuations
(described by the exponent $\theta$). Even the size need not be
gaussian like (i.e., $\nu\neq 1/2$).  We discuss various ways of
getting these quantities, some of which are general and some are very
particular for the directed polymer problem. 

\subsection{Replica}
\label{sec:replica}
Even in the absence of any knowledge of the probability distribution
like e.g. $P(\ln Z)$, various information can be extracted from the
moments.  This starts from the simple identity that
\begin{equation}
\label{eq:rep.1}
\dav{Z^n} = \dav{e^{n\ln Z}} = \exp \left( \sum 
\frac{1}{m!}n^m \dav[(c)]{(\ln Z)^m}\right ),
\end{equation}
where $\dav[(c)]{(\ln Z)^m}$ are the cumulants.  
As an interacting $n$ polymer problem, each polymer of length $N$, one
expects an extensive term and a correction which depends on $n$ and
$N$, as
\begin{equation}
\label{eq:rep.2}
\ln \dav{Z^n} = nN  \epsilon + {\tilde{\cal F}}(n,N).
\end{equation}
It is the correction term that is more useful and informative.
Our interest is generally in small $n$ and $N\rightarrow\infty$, as
shown in Fig. \ref{fig:nN}.  This also becomes clear from a more
familiar form for the free energy.  Eq. \ref{eq:rep.1} can be written
as
\begin{equation}
\label{eq:repl}
  \dav{\ln Z} =\lim_{n\rightarrow 0} \frac{\dav{Z^n} - 1}{n}.
\end{equation}
so that to compute the average free energy we may consider a case of
$n$-replicas of the original system or after averaging, an $n$-polymer
problem with extra interactions induced by the disorder though an
$n\rightarrow 0$ limit is to be taken at the end.  A few possible paths to
take the limit for long chains are shown in Fig. \ref{fig:nN}.

\fignN

Nontrivial results are expected if and only if the origin in Fig.
\ref{fig:nN} is a singular point so that the limits
$n\rightarrow 0$ and $N\rightarrow\infty$ become non-interchangeable.
In other words, the $n$ and $N$ dependences should be coupled so that
the appropriate path is a scaling path like (c) in the figure

\begin{exercise}
  {\bf Path dependence with rare events}\\
  (i) Consider the probability distribution of Eqs.
  \ref{eq:rar1},\ref{eq:rar2} and calculate $\dav{\ln x}$.\\
  (ii) Calculate the moments $\dav{x^n}$ and use Eq.  (\ref{eq:repl}).
  Go along paths (a) and (c) of Fig. \ref{fig:nN} and
  compare with  the result of (i). \\
  (iii) Show that correct average is obtained if a scaling path is
  chosen with $n\sqrt{N}=$ constant.
\end{exercise}

\subsubsection{$n\rightarrow 0$ and $N\rightarrow\infty$}
\label{sec:nright-0-nright}
Now, $\dav[(c)]{(\ln Z)^2}$ is related to the free energy fluctuation
and, as per Eq. (\ref{eq:7}), it is expected to scale as
$N^{2\theta}$.  If higher order fluctuations (or cumulants) do not
require any new exponent, then it is fair to expect $\dav[(c)]{(\ln
  Z)^m}\sim N^{m\theta}$.  A natural choice is 
${\tilde{\cal  F}}(n,N) = {\cal F}(nN^{\theta})$.

If we demand that $\ln \dav{Z^n}$ is proportional to $N$ for large
$N$, then, for $x=nN^{\theta}\rightarrow\infty$, ${\cal F}(x) \sim
x^{1/\theta}$ so that
\begin{equation}
\label{eq:rep.3}
\ln \dav{Z^n} = n \epsilon N + {\sf a} n^{1/\theta} N. \ (N\rightarrow\infty)
\end{equation}
This is for path (a) of Fig. \ref{fig:nN}.
In contrast, if we take $n\rightarrow 0$ for finite $N$,  path (c),  a
Taylor series expansion gives
\begin{equation}
\label{eq:pathc}
\ln \dav{Z^n} = n N\epsilon + {\sf a} n^2 N^{2\theta} + ... .\quad
(n\rightarrow 0)
\end{equation}
Eqs. \ref{eq:rep.3},\ref{eq:pathc} can be used to calculate $\theta$,
the free energy fluctuation exponent..

\begin{exercise}
Prove that $\ln \dav{Z^n}\propto N$. 
\end{exercise}

The size of a polymer can also be handled in this replica approach by
a careful limit as follows,
\begin{eqnarray}
%\lefteqn
&&{\dav{<r_N^2>} }%\nonumber\\
%&=& 
=\left [ \frac{\int{\cal DR}r_N^2 \exp(-\beta H)}{\int{\cal
DR} \exp(-\beta H)} \right ]_{\rm av}\nonumber\\
& =& \lim_{n\rightarrow 0}\left [ \left({\textstyle \int}{\cal
DR} \exp(-\beta H)\right)^{n-1}%\times\nonumber\\
%&&\qquad 
{\textstyle\int}{\cal DR} r_N^2 \exp(-\beta H)\right ]_{\rm av}.\nonumber\\
\end{eqnarray}
Using the identity $\lim_{n\rightarrow 0}\dav{ \{\int {\cal
    DR}\exp(-\beta H)\}^n}=1$, one may consider the partition function
of the $n$-replica problem, compute the value of average of $r_N^2$
(without normalization) for one of the replicas and then take
$n\rightarrow 0$.  However this procedure is not easy to implement.

\subsection{Bethe ansatz}
The replica approach requires an evaluation of $\dav{Z^n}$.  For a
gaussian distributed, delta-correlated disorder, $\dav{Z^{n}}$
corresponds to the partition function of an $n$-polymer system with
the Hamiltonian given by Eq. (\ref{eq:5}).  Noting the similarity with
the quantum Hamiltonian with $z$ playing the role of imaginary time,
finding $N^{-1} \ln\dav{Z^n}$ for $N\rightarrow\infty$ is equivalent
to finding the ground state energy $E$ of a quantum system of $n$
particles.  This problem can be solved exactly only in one dimension
($d=1$) using the Bethe ansatz.  This gives the ground state energy as
\begin{equation}
\label{eq:beth.2}
E= - K(n  - n^3 )\qquad (d=1)
\end{equation} 
which gives
\begin{equation}
\label{eq:beth.3}
\theta=\frac{1}{3} \Longrightarrow 
\nu=\frac{2}{3}. \ (d=1)\quad {\rm (see \ below)}
\end{equation}
The polymer has swollen far beyond the random walk or gaussian
behaviour.  What looks surprising in this approach is that there is no
``variance'' (2nd cumulant) contribution.  It is just not possible to
have a probability distribution whose variance vanishes identically!
This is a conspiracy of the $N\rightarrow \infty$ limit inherent in
the quantum mapping and the value of the exponent $\theta$ that
suppressed the second cumulant contribution (see Eq. (\ref{eq:rep.3}).
To repeat, in this quantum ground state approach, $N\rightarrow\infty$
has been taken first and so there are no higher order terms.

\begin{exercise}
Why cannot the Bethe ansatz be extended to higher dimensions?
\end{exercise}

\subsubsection{Flory approach}
Using the quantum analogy, we may try to estimate the ground state
energy in a simple minded calculation.  The elastic energy is like the
kinetic energy of quantum particles which try to delocalize the
polymers (random walk) while the attractive potential tries to keep
the polymer together.  For $n$ polymers there are $n(n-1)/2$
interactions.  We take the large $n$ limit so that if the particles
are bound in a region of size $R$, the energy is (using dimensionally
correct form with $R$ as the only length scale)
\begin{equation}
\label{eq:beth.4}
E= \frac{n}{R^2} - \frac{n^2\Delta}{R}
\end{equation}  
which on minimization gives $E\sim n^3$ consistent with the Bethe
ansatz solution.  At this point we see the problem of the replica
approach if the limit is taken too soon.  Since our interest is
eventually in $n \rightarrow 0$, we could have used in this argument
the linear term of the combinatorics.  That would have made energy
``extensive'' with respect to the number of particle and replaced the
disorder-induced attraction by a repulsion (note the negative sign).
The end result would however have no $n^3$ dependence.  This is a real
danger and any replica calculation has to watch out of these pitfalls.
Quite strangely we see that the correct answer came by taking
$n\rightarrow\infty$ first and then $n\rightarrow 0$ or, probably
better to say, by staying along the ``attractive part'' of the
interaction only.

\begin{exercise}
  Does the Bethe ansatz require $n\rightarrow\infty$?
\end{exercise}

\section{Can we handle free energy?}
\label{sec:can-we-handle}
So far we have been looking at the moments of the partition function
and trying to extract relevant information from that.  Is it possible
to avoid the replica trick altogether and treat the free energy
explicitly?

This is something unique to directed polymers that there is a way to
study the average free energy and implement RG directly for the free
energy bypassing the $n\rightarrow 0$ problem of Fig \ref{fig:nN}
completely.  This is a very important step because it gives an
independent way of checking the results of replica approach.
 
\subsection{Free energy and the KPZ equation }
For a polymer, the partition function satisfies a diffusion or
Schrodinger-like equation. This equation can be transformed to an
equation for the free energy $F({\bf r},z)=-T \ln Z({\bf r},z)$.  This is the free
energy of a polymer whose end point at $z$ is fixed at ${\bf r}$.  To
maintain the distance fixed at ${\bf r}$ a force is required which is
given by ${\bf g}=-\nabla F$.  If we want to increase the length of a
polymer by one unit, we need to release the constraint at the previous
layer (think of a lattice). The change in free energy would then
depend on the force at that point, and of course the random energy at
the new occupied site. The change $\partial F({\bf r})/\partial z$
being a scalar can then depend only on the two scalars $\nabla \cdot
{\bf g}$, $g^2$.  A direct derivation of the differential equation for
the free energy shows that these are the three terms required.  The
differential equation, now known as the Kardar-Parisi-Zhang equation,
is
\begin{equation}
\frac{\partial F}{\partial z} = \frac{T}{2K} \nabla^2 F - \frac{1}{2K}
(\nabla F)^2 + \eta({\bf r},z). 
\end{equation}
If we can solve this \textit{exact} equation and average over the random energy
$\eta$, we get all the results we want. 

One may also write down the equation for the force in this ``fixed
distance'' ensemble as
\begin{equation}
  \label{eq:12}
\frac{\partial {\bf g}}{\partial z} =  \frac{T}{2K} \nabla^2  {\bf g}
             -  \frac{1}{2K}  {\bf g}\cdot\nabla {\bf g} + \nabla \eta({\bf r},z).   
\end{equation}
This equation is known as the Burgers equation.

\subsection{Thermal correlation and an  exact result}
In the previous section we considered the force required to maintain
the distance fixed.  We now consider a fixed force ensemble that helps
us in determining the response functions.

It is known in statistical mechanics that the response of a system in
equilibrium is determined by the fluctuations.  If we consider the
response of a directed polymer to an applied force, the averaged
response function comes from the Hamiltonian
\begin{eqnarray}
\label{eq:2.3.1}
H&=& \frac{d}{2}K \int dz \left( \frac{\partial
{\bf r}}{\partial z}\right ) ^{^2} + \int \eta({\bf r}(z),z) - {\bf
g}\cdot \int \frac{\partial 
{\bf r}}{\partial z} \ dz \nonumber\\
&=&\frac{d}{2} K\int dz \left( \frac{\partial}{\partial z} ({\bf r} -
\frac{{\bf g}z}{dK}) \right ) ^{^{\scriptstyle 2}} + \int \eta({\bf
r}(z),z) -\nonumber\\
&&\quad\frac{1}{2} \frac{g^2 N}{d.K},
\end{eqnarray}
The disorder is gaussian-distributed as in Eq. (\ref{eq:3}).

The general response function for the force is
\begin{equation}
\label{eq:2.3.2}
 \left . C_T\right|_{_{ij}} = \left.\frac{\partial^2 \dav{\ln
Z}}{\partial g_i\partial g_j}\right |_{_{g=0}} = \dav{\langle r_i
r_j\rangle - \langle r_i\rangle \langle r_j\rangle},
\end{equation}
$i,j$ representing the components.

\subsubsection{Exact result on response: pure like}
\label{sec:exact-results}
By a redefinition of the variables and using the $\delta$-correlation
of the disorder in the $z$ direction, we have
\begin{equation}
\label{eq:2.3.3}
\dav{\ln Z({\bf g})} = \dav{\ln Z({\bf g}=0)} + \frac{g^2 N}{d.K}, 
\end{equation}
from which it follows that
\begin{equation}
\label{eq:2.3.4}
 C_T = \ \frac{TN}{d.K},
\end{equation}
as one would expect in a pure system, Eq. (\ref{eq:1.2}).  And there
are no higher order correlations.

Two things played important roles in getting this pure-like result: (i)
The disorder correlation has a statistical translational invariance
coming from the delta function in the $z$-coordinate, and (ii) the
quadratic nature of the hamiltonian.  If disorder had any correlation
along the length of the polymer, Eq. \ref{eq:2.3.4} will not be valid.

\begin{exercise}
Put a semi-flexibility term $(\partial^2 {\bf r}/\partial z^2)^2$
in the Hamiltonian and then derive the expression for $C_T$. 
\end{exercise}
\begin{exercise}
  Is Eq. \ref{eq:2.3.4} valid for a lattice problem?  Note also that
  the full range $[-\infty,\infty]$ has been used.
\end{exercise}
\begin{exercise}
Prove that 
\begin{equation}
  \label{eq:13}
 P(F({\bf r}_N)) = P(F(0) + \frac{d.K r_N^2}{2N}) 
\end{equation}
  Hint: Choose ${\bf g}= {d.K \bf r}/N$.
\end{exercise}

\subsection{ Free energy of extension: pure like}
\label{sec:free-energy-extens}
We want to know the free energy cost in pulling a polymer of length
$N$ from origin (where the other end is fixed) to a position ${\bf r}$.
For the pure case, the free energy follows from Eq. (\ref{eq:1.2})
(with $K$ inserted) as
\begin{equation}
\label{eq:1.4}
F({\bf r},N)- F({\bf o},N) = \frac{d}{2} \ K \ \frac{ r^2}{N}. \ ({\rm
  pure})
\end{equation}

For the disordered case, we use Eq. (\ref{eq:13}) to see the
surprising result,
\begin{equation}
  \label{eq:14}
\dav{F({\bf r},N)- F({\bf o},N)} = \frac{d}{2} \ K \ \frac{ r^2}{N}.\ ({\rm
 disorder})
\end{equation}
Therefore, on the average the stretching of a chain is pure-like
(elastic) with the same elastic constant though the fluctuation is
anomalous ($\theta \neq 0$).  This result has a far reaching
consequence that in a renormalization group procedure, the elastic
constant must remain an invariant.  As we shall see, this invariance
condition puts a constraint on $\nu$ and $\theta$, making only one
independent.

\subsection{RG of the KPZ equation}
To analyze the nonlinear KPZ equation, an RG procedure may be adopted.
This RG is based on treating the nonlinear term in an iterative manner
by starting from the linear equation.  This is a bit unusual because
here we are not starting with a ``gaussian'' polymer problem, rather,
a formal linear equation that \textit{does not} satisfy the exponent
relation Eq. (\ref{eq:15}).  Leaving aside such peculiarity, one may
implement the coarse-graining of RG to see how the couplings change
with length scale.

\subsubsection{Scale transformation and an important relation}
\label{sec:scale-transf-an}
As a first step, we do the scale transformation
\begin{equation}
x\rightarrow bx, z\rightarrow b^{1/\nu} z, 
\ {\rm and}\  F \rightarrow b^{\theta/\nu}F, 
\end{equation}
so that the randomness transforms like
\begin{eqnarray}
\lefteqn{\dav{\eta({\bf r}_1,z_1) \eta({\bf r}_2,z_2)}}\hspace{1cm}\nonumber\\
& =& \Delta \delta({\bf
r}_1-{\bf r}_2) \delta(z_1-z_2) \nonumber\\
&\rightarrow& b^{-d-(1/\nu)} \Delta  \delta({\bf
r}_1-{\bf r}_2) \delta(z_1-z_2)
\end{eqnarray}
This transformation done on Eq. (\ref{eq:14}) shows that for $K$ to be
an invariant (no $b$-dependence) we must have,
\begin{equation}
  \label{eq:15}
  \fbox{$\theta+1 = 2\nu$}
\end{equation}
This is trivially valid for the gaussian pure polymer problem but
gives a relation between the free energy fluctuation and the size of
the polymer.  This is borne out by the intuitive picture we develop
below.  This relation gives the size exponent $\nu=2/3$ in $d=1$ (See
Eq.  (\ref{eq:beth.3})).

The equation in terms of the transformed variables is then
\begin{eqnarray}
\frac{\partial F}{\partial z} &=& \frac{T}{2K}b^{(1-2\nu)/\nu} \nabla^2
F - \frac{1}{2K} b^{(\theta-2\nu+1)/\nu} (\nabla F)^2 +\nonumber\\
&&\qquad b^{(1-d\nu-2\theta)/\nu} \eta({\bf r},z)
\end{eqnarray}
The $b$-dependent factors can be absorbed  to define new parameters, except
for $K$.

The temperature however gets renormalized.  Its flow is described
by the flow equation
  \begin{equation}
\label{eq:1.2.2}
L\frac{\partial T}{\partial L} = \frac{1-2\nu }{\nu} T  \ ({\rm to \
  leading \ order})
\end{equation}
For $\nu>1/2$, $T(L) \rightarrow 0$.  The disorder dominated phase is
therefore equivalent to a zero temperature problem.  In other words,
the fluctuation in the ground state energy and configurations dominate
the behaviour at finite temperatures in situations with $\nu>1/2$.  It
is this renormalization that was missing in Sec \ref{sec:zrg}.

The nonlinearity (or $g^2$) contributes further to the renormalization
of the temperature through the appropriate dimensionless variable
$u=(K\Delta/T^3) L^{2-d}$ (same as in Eq. (\ref{eq:1}).  The important
flow equation is for this parameter $u$ (upto constant factors) and it is
\begin{equation}
  \label{eq:21}
L\frac{du}{dL} = (2-d)u + \frac{2d-3}{4d} u^2.
\end{equation}
One immediately sees a major difference with the flow equation Eq.
(\ref{eq:uflow}) for $d=1$.  Because of the change of sign of the
quadratic term in Eq. \ref{eq:21}, there is now a f.p. for $d=1$ as
shown in Fig. \ref{fig:fp}(b).  This flow equation does not behave
properly in a range $d\in [1.5,2)$ but that is more of a problem of RG
than DP as such and so, may be ignored here.  Note also that no extra
information can be obtained for $d\ge 2$ other than what we have
obtained so far in Sec. \ref{sec:disorder-relevant}.  A point to
stress is that the fixed point at $d=1$ gives back the exact exponents
we have derived so far.

\section{Virtual reality: Intuitive picture}
\label{sec:intuitive-picture}

Powered by the quantitative estimates of the free energy fluctuation
and size exponents, we now try to generate a physical picture.  To do
this we exploit a few more exact results.

\subsection{Rare events}
\label{sec:intu-pict-rare}
How to understand the exact result of Eq. (\ref{eq:2.3.4})?  We have
shown that there is a low temperature region or in lower dimensions,
disorder or randomness results in a new phase but then its response is
the same as the pure system?

For the pure case as $N\rightarrow\infty$ the width of $P({\bf
  r}_N,N)$ increases.  Hence the increase of $C_T$ with $N$.  With
randomness, for $T\rightarrow 0$ we need to look for the minimum
energy path.  Let us suppose that there is a unique ground state, i.e.
$E({\bf r}_N)$ or $F({\bf r}_N)$ is a minimum for a particular path.
This tells us that as the temperature is changed $T$ still low, the
polymer explores the nearby region so that the probability
distribution gains some width which is determined by the thermal
length.  Susceptibility would be the width of the distribution and
this is independent of $N$.  This cannot satisfy the relation. Since
the relation is valid on the average, one way to satisfy it is to
assume that most of the samples have a unique ground-state but once in
a while (rare samples) there may be more than one ground state which
happens to be far away from each other.  Suppose there are such rare
samples, whose probabilities decay as $N^{-\kappa}$, where the paths
are separated by $N^{\nu}$, then the contribution to the fluctuation
from these samples would be $N^{2\nu -\kappa}$.  In case $\kappa=2\nu
-1$, we get back the exact result.  The relation of Eq. (\ref{eq:15})
tells us $\kappa=\theta$.  The rare events control the free energy
fluctuation.

What we see here that though the average behaviour is the same as that
of the pure system, the underlying phenomenon is completely different;
the average thermal response is determined by the rare samples that
have widely separated degenerate ground state and the probability of
such states also decay as a power law of the chain length.

In a given sample elastic energy $\sim {r^2}/N \sim N^{2\nu -1}$.  The
pinning energy would also grow with length say as
$N^{\tilde{\theta}}$.  We see, $\tilde{\theta}=2\nu -1=\theta$.  One
way to say this is that the sample to sample fluctuation and the
energy scale for a given sample are the same.

\figriv

These results can now be combined for an image of the minimum energy
paths.  If the end points at $z=N$ are separated by $r \ll N^{\nu}$,
the paths remain separated (each path exploring an independent
disordered region) until they join at $\Delta z\sim r^{1/\nu}$, after
which they follow the same path.  If the end points are separated by a
distance $r \gg N^{\nu}$, the two paths explore independent regions
and they need not meet.  This (Fig.\ref{fig:river}) picture is often
alluded to as the river-basin network.

\subsection{Probability distribution}
For a pure polymer, the probability distribution of the end point is
gaussian but it need not be so for the disordered case.  One way to
explore the probability distribution is to study the response of the
polymer as we take it out of its optimal or average position, e.g. by
applying a force.  In a previous section we used the fixed distance
ensemble where the end point was kept fixed and we looked at the force
$g$ required to maintain that distance (see Eq. (\ref{eq:12})). Here
we consider the conjugate fixed force ensemble.

\subsubsection{Response to a force}
\label{sec:response-force}
Let us apply a force that tries to pull the end of the polymer beyond
the equilibrium value $r \sim R_0$.  In equilibrium, the average size
$R$ or extension by the force can be expressed in a scaling form
\begin{equation}
\label{eq:1.6}
R= R_0 \ {\sf f}\left(\frac{gR_0}{ T}\right).
\end{equation}
This is because for zero force one should get back the unperturbed
size while the force term may enter only in a dimensionless form in
the above equation where the quantities available are the size $R_0$
and the thermal energy.   For small $g$,
linearity in $g$ is expected. This requirement gives
\begin{equation}
\label{eq:1.7}
R= R_0 \frac{R_0}{T} g \quad (k_B=1),
\end{equation}
$R$ is not proportional to $N$ if $\nu \neq 1/2$ ($R_0 \sim N^{\nu}$).
The polymer acts as a spring with $T/R_0^2$ as the effective spring
constant.

\subsubsection{Scaling approach}
\label{sec:scaling-approach}
Let us develop a physical picture and the corresponding algebraic
description (called a scaling theory).  The polymer in absence of any
force has some shape of characteristic size $R_0$.  The force
stretches it in a way that it breaks up into blobs of size
$\xi_g=T/g$.  For size $< \xi_g$ the polymer looks like a chain
without any force and these blobs, connected linearly by geometry, act
as a ``new'' polymer to respond to the force by aligning along it.  We
now have two scales $R_0$ and $\xi_g = T/g$.  A dimensionless form is
then
\begin{equation}
\label{eq:1.8}
R= R_0 {\sf f}\left(\frac{R_0}{\xi_g}\right) \sim  \frac{R_0^2}{T} 
\end{equation}
Now each blob is of length $N_g$ so that $\xi_g= N_g^{\nu}$ and there
are $N/N_g$ blobs.  We therefore expect 
\begin{equation}
\label{eq:1.8a}
R= \frac{N}{N_g} \xi_g = N \xi_g^{1-1/\nu} = N \left
( \frac{g}{T}\right)^{^{(\nu-1)/\nu}}
\end{equation}
This gives a susceptibility $\chi = \partial R/ \partial g \sim
g^{(1-2\nu)/\nu}$ .

\subsubsection{And so the probability distribution is}
\label{sec:so-prob-distr}
Let us try to get the susceptibility of Eq. (\ref{eq:1.8a}) in another way.  Let us assume
that the probability distribution for large $R$ is
\begin{equation}
  \label{eq:16}
  P(R) \sim \exp\left(- ({r}/{R_0})^{\delta}\right).
\end{equation}
The entropy is given by $S(r) = -\ln P$.  The free energy in
presence of a force which stretches the polymer to the tail region is
given by
\begin{equation}
  \label{eq:17}
 F=T({r}/{R_0})^{\delta} - g r. 
\end{equation}
This on minimization gives $g= \frac{T}{R_0} ({r}/{R_0})^{\delta-1}$.
By equating this form with Eq. (\ref{eq:1.8a}), we get
$\delta={1}/{(1-\nu)}$ and
\begin{equation}
\label{eq:1.9}
P(R) \sim \exp\left[-\left(\frac{r}{R_0}\right)^{{1/(1-\nu)}}\right].
\end{equation}
For $\nu=1/2$ we do get back the gaussian distribution. 

\figblob

The above analysis, done routinely for polymers, relies on the fact that
there is only one length scale in the problem, namely, the size of the
polymer.  If we are entitled to do the same for the disorder problem,
namely only one scale, $R_0 \sim N^{\nu}$, matters, then the blob
picture goes through in toto.  The chain breaks up into ``blobs'' and
the blobs align as dictated by the force.  Each blob is independent
and the polymer inside a blob is exploring its environment like a DP
pinned at one end.  The probability distribution is therefore given by
Eq. \ref{eq:1.9} which for $d=1$ is
\begin{equation}
  \label{eq:18}
 P(r) \sim \exp(-|r|^3/t^2)  \quad (d=1).
\end{equation}
If we use the relation $\Delta F\equiv F(x,N)-F(0,N)\sim x^2/N$, then
the above probability distribution can be mapped to the distribution
of the free energy as
\begin{eqnarray}
  \label{eq:19}
  P(F)  &\sim&
  \exp\left[-\left(\frac{|\Delta
        F|}{N^{\theta}}\right)^{1/2(1-\nu)}\right]\nonumber\\
&\sim& \exp \left(- \frac{|\Delta F|^{3/2}}{N^{1/3}}\right ) \quad (d=1).
\end{eqnarray}

\subsection{Confinement energy}
Suppose we confine the polymer in a tube of diameter $D$.  This is
like the localization length/argument used to justify the energy in
the quantum formulation.  The polymer in the random medium won't feel
the wall until its size is comparable to that, $D\sim N_0^{\nu}$ which
gives the length at which the polymer feels the wall.
Elastic energy of a blob is $D^2/N_0$.  But
because of the tube, the polymer will be stretched in the tube
direction.  The number of blobs is $N/N_0$, so that the energy is 
\begin{equation}
\frac{N}{N_0} \ \frac{D^2}{N_0} = N \ 
\left(\frac{D}{N_0}\right )^{^{\scriptstyle 2}}
= N\ \frac{1}{D^{2(1-\nu)/\nu}}.
\end{equation}
This gives the known form $1/D^2$ used in the quantum analogy
(consistent with dimensional analysis) but for $\nu=2/3$, this gives
$1/D$.

A cross-check of this comes from the energy of a blob. Each blob has the
fluctuation energy $N_0^{\nu}$ and so free energy per unit length
$F/N\sim N_0^{\theta}/N_0\sim D^{-2(1-\nu)/\nu}$.

\section{Overlap}

The problem we face in a disordered system is that there is no well
defined ground state - the ground state is sample dependent.  There is
therefore no predefined external field that will force the system into
its ground state (e.g., a magnetic field in a ferromagnetic problem).
This is a generic problem for any random system.

But, suppose, we put in an extra fictitious (ghost) polymer and let it
choose the best path.  Now we put in the actual polymer in the same
random medium but with a weak attraction $v$ with the ghost. At $T=0$,
this DP will then sit on top of the ghost.  In absence of any
interaction ($v\rightarrow 0$ ), the DP would go over the ghost in any
case if there is a unique ground state, not otherwise.  At non-zero
temperatures there will be high energy excursions and how close to the
ground state we are will be determined by the number of common points
of the two polymers.  This is called overlap which may be
quantitatively defined as
\begin{equation}
\label{eq:2.4.1}
q_i=\frac{1}{N} \int \delta({\bf r}_1(z)-{\bf r}_2(z)),
\end{equation}
for a given sample $i$ and then one has to average over the disorder
samples, $q\equiv\dav{q_i}$.

In case of a repulsive interaction, the situation will be different.
If there are more than one ground state, the two chains will occupy
two different paths and there is no energetic incentive to collapse on
top of each other when the repulsion $v\rightarrow 0+$.  In such a
scenario, the overlap $q(v\rightarrow 0+)\neq q(v\rightarrow 0-)$.
Conversely, a situation like this for the overlap would indicate
presence of degenerate ground states.  For a unique ground state, the
second chain would follow a nearby excited path with certain amount of
overlaps with the ground state.

In a replica approach, such occupancy of different ground states may
be achieved by ``replica symmetry breaking'' (i.e. various replicas
occupying various states) but the difficulty arises from the
$n\rightarrow0$ limit.  In the case of directed polymer, we have
argued that the degenerate states occur only rarely and therefore the
effect of ``replica symmetry breaking'' if any has to be very small.
This is why the Bethe ansatz gave correct results without any breaking
of replica symmetry.

The method to compute the overlap was developed by Mukherji.  By
introducing a repulsive potential $v\int dz \delta({\bf r}_1(z)-{\bf
  r}_2(z))$ for the two polymers in the same random medium, the free
energy $F({\bf r}_1(z),{\bf r}_2(z),z,v)$ can be shown to satisfy a
modified KPZ type equation
\begin{equation}
  \label{eq:20}
\frac{\partial F}{\partial z} = \sum_i \left (\frac{T}{2K} 
\nabla^2_i F - \frac{1}{2K}
(\nabla_i F)^2 + \eta({\bf r}_i,z)\right)+ v\delta({\bf r}_1-{\bf r}_2).   
\end{equation}
which can be studied by RG.  The mutual interaction has no effect on
the single chain behaviour but the interaction gets renormalized.  The
flow equation for the dimensionless parameter $u$ of Eq. \ref{eq:21}
remains the same.  The exponent relation of Eq. \ref{eq:15} also
remains valid.  The interaction gets renormalized as
\begin{equation}
  \label{eq:22}
  L\frac{\partial v}{\partial L}= \left(\frac{1-\theta}{\nu} -d +
    \frac{u}{2}\right) v ,
\end{equation}
where $v$ is in a dimensionless form.  For the pure problem
($\theta=0,\nu=1/2$) this reduces to the expected flow equation of Eq.
(\ref{eq:uflow}) for repulsive interaction ($ u \rightarrow -u$). For
overlap one needs only the first order term because we need
$v\rightarrow 0$.
 
The overlap can be written in a polymer-type scaling form $q
=N^{\Sigma}\  {\sf f}(v N^{-\phi\nu})$, where $\Sigma=\theta- \phi\nu
-1$.  The above RG equation for $v$ shows that the exponent $\Sigma
=0$ at the stable fixed point for $u$ of Fig \ref{fig:fp}(b) at $d=1$.
However, $\Sigma< 0$ at the transition point for $d>2$.  This means
that the overlap vanishes at the transition point from the strong
disorder side as $q\sim |T-T_c|^{|\Sigma|\zeta}$.

This approach to overlap  can be extended to $m$-chain overlaps also, which
show a nonlinear dependence on $m$ at the transition point.  This
suggests that eventhough the size exponent is $\nu=1/2$ gaussian like,
there is more intricate structure than the pure gaussian chain.

\section{Summary and open problems}
\label{sec:summ-open-probl}
The behaviour of a directed polymer in a random medium in 1+1
dimensions seems to be well understood.  There is a strong disordered
phase at all temperatures.  For $d>2$ renormalization group analysis
shows a phase transition from a low temperature strong disordered
phase to a weak disorder, pure-like phase.  There are rare
configurations with degenerate widely separated ground states, giving
a contribution to ``overlap''.  However most questions on the higher
dimensional strong-disorder phase remain open.
\acknowledgments
It was a pleasure to give this set of lectures at the SERC school held
at IMSC, Chennai, in April 2003.

\end{document}